\newif\ifTwoColumn%
\newif\ifSUBMIT%
\newif\ifCOMMENTS%
\newif\ifFIGs%
\newif\ifFIGoneColumn%
\let\ifSUBMIT\iftrue%
\let\ifCOMMENTS\iftrue%
\let\ifFIGoneColumn\iftrue%
\documentclass[journal=langd5]{achemso}
\setkeys{acs}{articletitle = true}
\setkeys{acs}{doi = true}
\mciteErrorOnUnknownfalse
\setkeys{acs}{etalmode = truncate, maxauthors = 1000}

\SectionNumbersOn
\usepackage{hyperref}
\usepackage{graphicx}
\usepackage{amsmath}
\usepackage{xcolor}
\usepackage{todonotes}
\usepackage{menukeys}
\usepackage{acronym}
\usepackage{array,mathtools,amssymb,booktabs}
\usepackage{physics}
\usepackage{placeins}
\usepackage{multirow}

\ifSUBMIT%
  \ifCOMMENTS%
    \usepackage{color}    
    \usepackage[normalem]{ulem}

    \def\STRIKE#1{{\color{red}\sout{#1}}}
    
    \def\NSTRIKE#1{{\color{red}\sout{#1}}}
  \else

    \def\STRIKE#1{}
    
    \def\NSTRIKE#1{}
  \fi
\else
  \usepackage{color}    
  \usepackage[normalem]{ulem}
 \definecolor{mygreen}{RGB}{0,180,0}    
  
  \def\STRIKE#1{{\color{red}\sout{#1}}}

  \def\NSTRIKE#1{{\color{blue}\sout{#1}}}
\fi

\usepackage{xcite}

\usepackage{xr}
\makeatletter
\newcommand*{\addFileDependency}[1]{
  \typeout{(#1)}
  \@addtofilelist{#1}
  \IfFileExists{#1}{}{\typeout{No file #1.}}
}
\makeatother
\newcommand*{\myexternaldocument}[1]{
    \externaldocument{#1}
    \addFileDependency{#1.tex}
    \addFileDependency{#1.aux}
}
\myexternaldocument{./paper_Pdot_SI}
\title[Pdot]%
{Binding Affinity between Polymer Dots (Pdots) and Ovalbumin Protein at Varying pH }

\author{Xingfei Wei}
\affiliation{Department of Chemistry, Johns Hopkins University, 
 Baltimore, Maryland 21218, USA}

\author{Wandi Xu}
\affiliation{Department of Chemistry, Johns Hopkins University, 
Baltimore, Maryland 21218, USA}

\author{Kushani Mendis}
\affiliation{Department of Chemistry and Biochemistry, University of Maryland, Baltimore County, 
Baltimore, Maryland 21250, USA}

\author{Zeev Rosenzweig}
\affiliation{Department of Chemistry and Biochemistry, University of Maryland, Baltimore County, 
Baltimore, Maryland 21250, USA}

\author{Rigoberto Hernandez}
\email{r.hernandez@jhu.edu}
\affiliation{Department of Chemistry, Johns Hopkins University, 
 Baltimore, Maryland 21218, USA}
\alsoaffiliation{Department of Chemical \& Biomolecular Engineering, 
Johns Hopkins University, Baltimore, Maryland 21218, USA}
\alsoaffiliation{Department of Materials Science \& Engineering, 
Johns Hopkins University, Baltimore, Maryland 21218, USA}


\date{\today}

\begin{document}
\newlength\figurewide
\ifFIGoneColumn
  \figurewide=.5\columnwidth
\else
  \figurewide=.9\columnwidth
\fi

\begin{tocentry} 
\ \\
   \includegraphics[clip=true,clip=true,height=5cm]{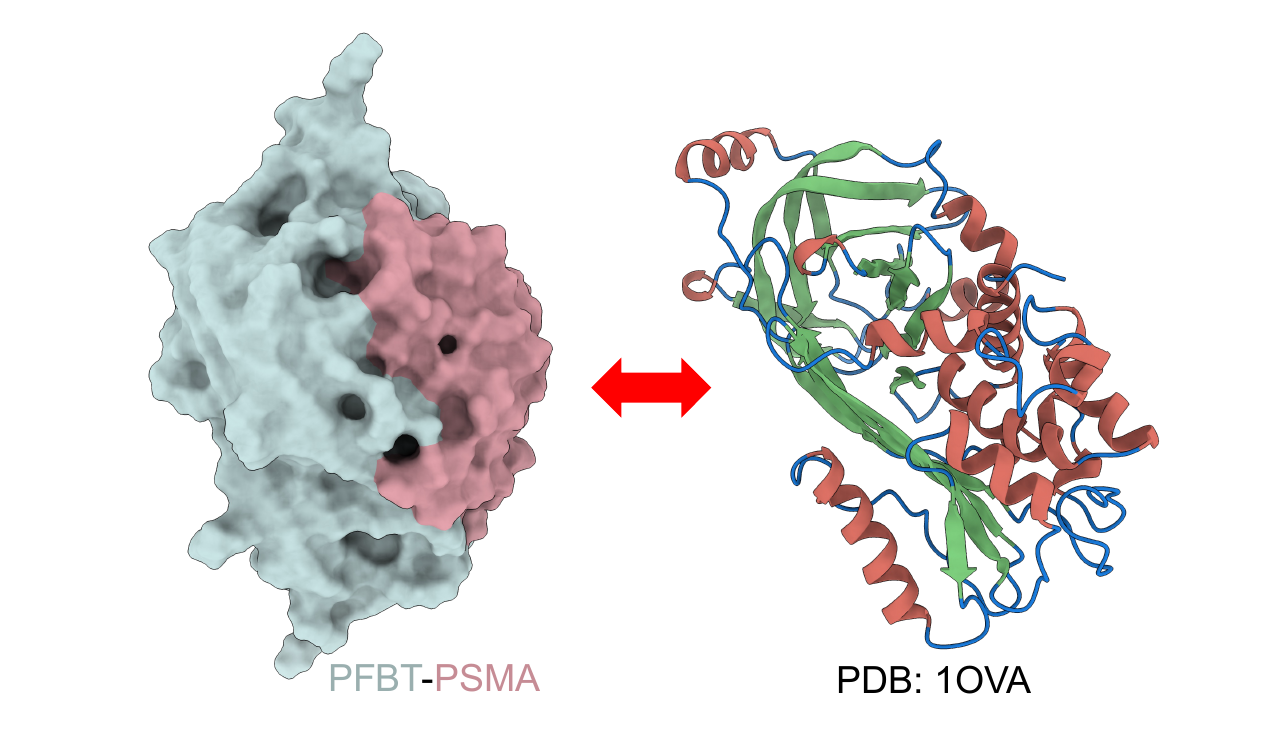}
\ \\
\end{tocentry}


\begin{abstract} 
Determining the binding mechanisms between polymer dots (Pdots) and proteins
is important for developing novel nanotechnologies for biomedicine and bioimaging.
In this work,
we use all-atom molecular dynamics (MD) simulations to determine
the binding affinity of Pdots with ovalbumin protein at pH = 7 and 1.
The selected Pdots are mixtures of
Poly[(9,9-dioctylfluorenyl-2,7-diyl)-alt-co-(1,4-benzo-(2,1$^\prime$,3)-thiadiazole)]
(PFBT)
and poly(styrene/maleic anhydride) (PSMA)
with varying composition.
At pH = 7, the Pdots have a net negative charge 
due to the COO$^-$ functional groups on the PFBT,
and the protein also has a net negative charge.
At pH = 1, the Pdots are charge neutral with
PFBT containing only COOH functional groups,
and the protein also has a net positive charge.
We sample the initial position of the protein by varying
its initial position through all 6 orientations of a cube.
For each orientation, we pull the
protein towards the PFBT region of the Pdot.
We compare the Coulombic and Lennard-Jones (LJ) interaction energies for the 6 different interacting faces and
two selected pH values.
We find that the LJ interaction energies are similar among all 
12 of these cases.
The measured Coulombic interaction energies suggest
that pH = 1 has better binding affinity than pH = 7.
The potentials of mean force (PMFs) along the pulling coordinates
differ with pH.
The PMFs from 2 of the 6 initial configurations
at pH = 1 are negative
whereas none of them are negative at pH = 7,
confirming the preferred binding affinity when pH = 1.
One of the faces at pH = 1 has the lowest PMF of $\sim-30$ kcal/mol,
which is much lower than the $\sim6$ kcal/mol seen
for the lowest case at pH = 7.
Comparison of protein residue charge distributions at pH = 7 and 1
further shows that the electrostatic interaction is critical to the binding affinity,
and negatively charged residues reduce at pH = 7 
does not bind to negatively charged Pdot.

\end{abstract}

\maketitle


\acrodef{PFBT}{poly[(9,9-dioctylfluorenyl-2,7-diyl)-alt-co-(1,4-benzo-(2,1$^\prime$,3)-thiadiazole)]}
\acrodef{PSMA}{poly(styrene/maleic anhydride)}
\acrodef{Pdot}{polymer dot}
\acrodef{MD}{molecular dynamics}
\acrodef{LAMMPS}{Large-scale Atomic Molecular Massively Parallel Simulator}
\acrodef{OPLS}{Optimized Potentials for Liquid Simulations}
\acrodef{COM}{center of mass}
\acrodef{LJ}{Lennard-Jones}
\acrodef{SI}{Supporting Information}\acused{SI}
\acrodef{PMF}{potential of mean force}
\acrodef{WHAM}{Weighted Histogram Analysis Method}
\acrodef{RMSD}{root-mean-square deviation}
\acrodef{Rg}[\protect$R_{\text{g}}$]{radius of gyration}
\acrodef{PAH}{poly(allylamine hydrochloride)}
\acrodef{DLVO}{Derjaguin-Landau-Verwey-Overbeek}
\acrodef{DPD}{dissipative particle dynamics}
\acrodef{vdW}{van der Waals}
\acrodef{FRET}{fluorescence resonance energy transfer}
\acrodef{ACCESS}{Advanced Cyberinfrastructure Coordination Ecosystem: Services \& Support}

\section{Introduction}
Determining the binding interaction between nanomaterials and biological systems
is important to developing new nanotechnologies for drug delivery,\cite{hern15d,jacob2018}
biomedicine,\cite{elsayed12a,hern15d}
diagnostics,\cite{pian2024,hern15d} and bioimaging.\cite{TXu2024,hern15d,hern22e,hern22j}
For example, the shape of nanoparticles has been reported
to affect nanoparticle endocytosis in living cells,
and their related nanoparticle toxicity.\cite{hern16j}
The RuBisCo protein in chloroplasts 
has better binding affinity
to positively charged Fe$_{3}$O$_4$ nanoparticles
than the negatively charged ones,
and this can affect RuBisCO protein corona formation 
and their function in plants.\cite{hern25d}
The hydrophobicity of nanoparticles is also directly related to their interaction 
with the biointerface in plant cells.
Specifically, Klaper and coworkers\cite{klaper26}
recently reported
that decreasing nanoparticle hydrophobicity can enhance the internalization
but the translocation of nanoparticles inside cells is not affected.

Theoretical methods and simulation models have been successfully
applied to compute the physical and chemical properties of nanomaterials,
and to cover the dynamics of nano-bio interactions.\cite{murphy18,hern19b,hern22d,hern24f,hern25g}
Combining experiments and \ac{MD} simulations,
the Murphy lab\cite{murphy18,hern18f}
demonstrated that
cationic nanoparticles, e.g. \ac{PAH} wrapped gold nanoparticle,
can interact with lipid vesicles and extract phospholipids
forming lipid corona on the nanoparticle surface.
The Hernandez lab\cite{hern22d,hern25b}
found that the zeta-potential of nanoparticles is critical 
to their binding affinities.
The latter was found approximately through
theoretical solutions of the Poisson-Boltzmann equation,
more accurately computed using \ac{MD} simulation,
and estimated using the \ac{DLVO} theory at high salt concentration.
Using $^1$H NMR spectra in solution, 
the surface curvature and ligand length were found
can affect the packing density,
ligand island structure, and headgroup mobility on nanoparticles;
this result is also demonstrated by All-Atom \ac{MD} simulations.\cite{hern19b,hern25g}
Using multiscale simulation methods---viz. All-Atom, MARTINI, and \ac{DPD} models---%
the Hernandez lab also reported the binding affinities of
different gold nanoparticles,
varying ligand types and nanoparticle size,
with Cytochrome C protein and lipid bilayer.\cite{hern19c,hern20g,hern24f,hern25h}

Conjugated polymer nanoparticles 
have received significant recent attention 
because they offer distinct advantages---lower cost and more flexibility---%
over inorganic ones
in many applications
involving light emitting diodes or chemical sensors.\cite{pecher2010}
Moreover,
revealing the protein corona formation mechanism
around polymer nanoparticles or \acp{Pdot} 
is very important to developments in nanomedicine,
drug delivery, and fluorescence imaging.\cite{nienhaus2023}
Recently, \citet{YLi2024} developed a new fluorescent \ac{Pdot} using \ac{PFBT}
which can perform
high-resolution imaging of intact meningeal blood vessels,
that is significant to neurology and neurosurgery.
The Rosenzweig lab \cite{zeev22} also developed
a similar near-infrared fluorescent \ac{Pdot} by mixing \ac{PFBT} and \ac{PSMA},
and it has been shown to be useful in plant cell imaging.
Using time-dependent \ac{FRET} spectroscopy,
\citet{zeev25} further demonstrated the dynamical process of
ovalbumin protein binding and corona formation on the \ac{Pdot} surface made of \ac{PFBT} and \ac{PSMA},
for varying protein concentration, salt concentration, and pH.
They found that reducing pH can enhance the binding affinity of \ac{Pdot} and ovalbumin protein.\cite{zeev25}
However,
the fundamental mechanism of 
how the electrostatic and \ac{vdW} interactions contribute to the binding affinity remains unclear.

In this work, we use All-Atom \ac{MD} simulations to elaborate the sensitivity
of the binding affinity
of \ac{Pdot} (made of \ac{PFBT} and \ac{PSMA}) with ovalbumin protein to pH.
As the isoelectric point (pI) of ovalbumin is 4.6, it is safe to assume that 
typical organismal conditions will be more basic.
We consider two extreme cases,
one (at pH = 7) which is strongly basic relative to the pI 
and one (at pH = 1) which is strongly acidic relative to the pI,
and model ovalbumin structure in each case accordingly.
The \ac{PFBT} part of the \ac{Pdot} contains COOH functional groups;
at pH = 7, the COOH groups are deprotonated becoming COO$^-$;
and at pH = 1, the COOH groups are charge neutral.
Meanwhile, the \ac{PSMA} part of the \ac{Pdot} stays charge neutral and inert varying pH.
We determined
the binding interactions of \ac{Pdot} starting
from 6 different faces on the protein,
by rotating the protein along the axes of a cube (like a dice);
see Figure~\ref{fig:model}.
The \ac{Pdot} surface is initially set $\sim2$ nm away from the protein surface.
We pull the \ac{Pdot} toward the protein
with the \ac{PFBT} part facing the protein, until the two surfaces bind.
The \ac{LJ} and electrostatic interaction energies
between the protein and Pdot
are tracked and recorded during the pulling process.
We also applied the umbrella sampling method to calculate the \ac{PMF}
by sampling states along the pulling trajectory.
The trends in the resulting values are compared across the 6 different initial
protein orientations,
and the two selected values of pH:  7 (neutral) and 1 (acidic).

\section{Model and Methods}
\label{sec:method}

\subsection{Simulation model}
\label{sec:model}
\begin{figure}[t]
\includegraphics[clip=true,scale=0.25,width=0.8\linewidth]{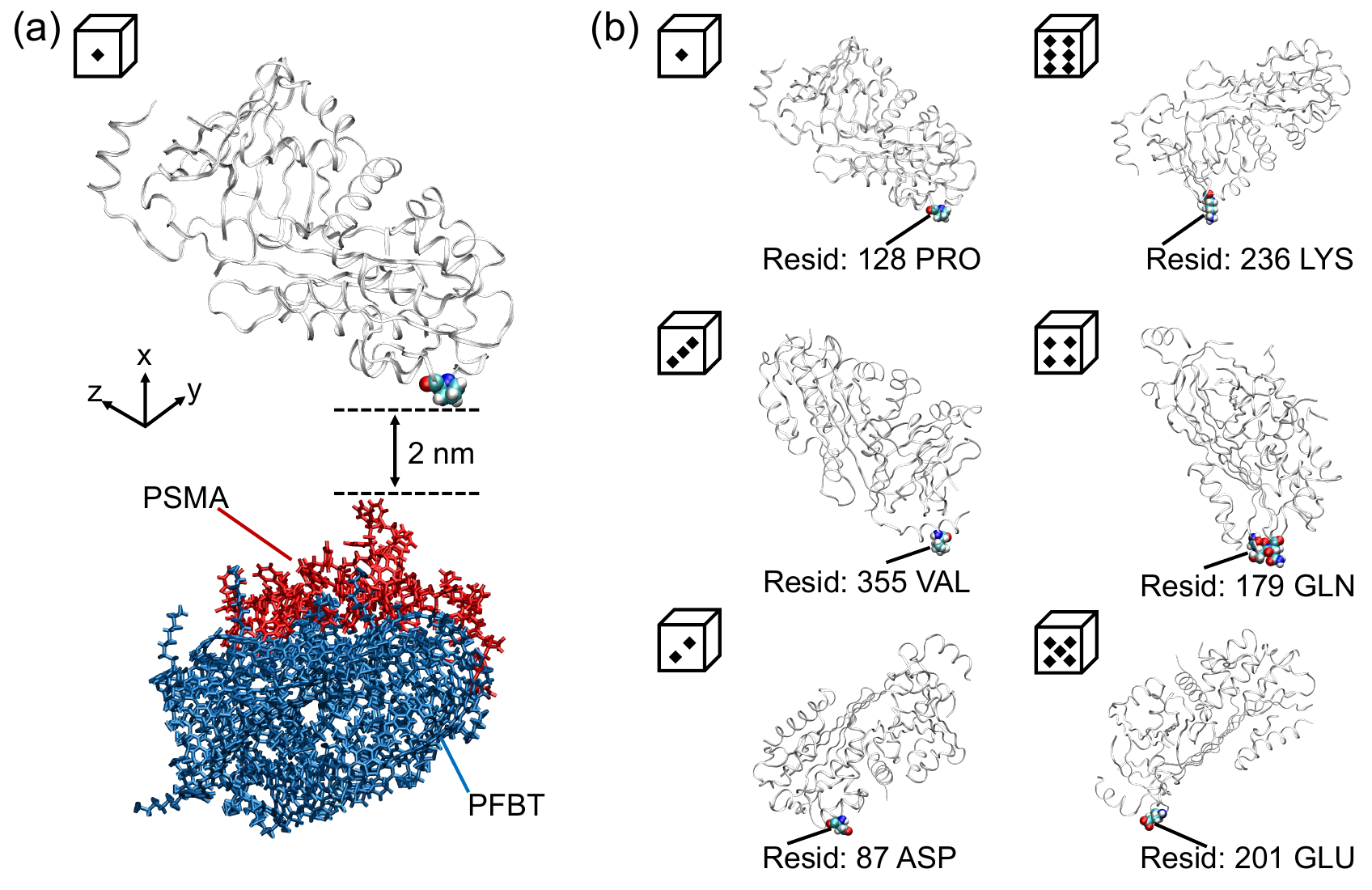}
\caption{
Schemes of the \ac{Pdot} interacting with the ovalbumin protein in 6 difference faces at pH = 7.
(a) Representative structure of the \ac{Pdot} interacting with the protein face 1,
and the distance between the protein surface and \ac{Pdot} surface is about 2 nm.
(b) Representative conformations of the 6 different interacting faces on the protein.
We use the dice faces 1 to 6 to label the 6 different interacting faces on the protein.
The corresponding schemes at pH = 1 is shown in Figure~S4 in the \ac{SI}.
}
\label{fig:model}
\end{figure}

Figure~\ref{fig:model} shows that the \ac{Pdot} model is made of \ac{PSMA}
and \ac{PFBT} wrapping together.\cite{zeev22,zeev25}
Figure~S1 in the \ac{SI} shows that
the \ac{PSMA} part is a polymer chain with 18 repeating units
(3 polystyrene blocks and 1 maleic anhydride block in each unit)
and the \ac{PFBT} is a polymer chain 44 repeating units.
The maleic anhydride block in \ac{PSMA} part of the \ac{Pdot}
contains COO$^-$ groups with negative charges at pH = 7.
We neutralize the system at pH = 7 by adding Na$^+$ ions.
At pH = 1, the maleic anhydride block is charge neutral with COOH groups
and we do not add any counter-ions.
The \ac{OPLS} force field \cite{jorgensen96, jorgensen01}
is used to simulate the atomistic interactions in polymers.
The \ac{Pdot} is relaxed by 2 steps shown in Figure~S2 in the \ac{SI}:
First, using implicit water model,
setting the dielectric constant to 80,
in a large box of 120 nm $\times$ 120 nm $\times$ 120 nm,
we make the linear polymers wrap into a \ac{Pdot}
by applying a steering force on the polymers towards the center of the box at (0,0,0)
under $NVT$ at 300 K.
Second, we reduce the box size to about 5 nm $\times$ 5 nm $\times$ 5 nm
and solvate the \ac{Pdot} system with TIP3P water molecules;
then we continue applying the steering force on the \ac{Pdot} under $NVT$ at 300 K for 5 ns;
then relax under $NPT$ at 1 atm and 300 K for 2 ns, and followed by $NVT$ at 300 K for 2 ns.
The steering force is 0.1 to 0.5 kcal/mol/{\AA} with cutoff radius of 2 nm.
In the end,
the \ac{Rg} of \ac{Pdot}---after relaxation runs---is about 2 nm.

Ovalbumin---the major protein in avian egg white, PDB structure 1OVA---%
is used in this work.\cite{zeev25}
Figure~S3 in the \ac{SI} shows that the ovalbumin protein has 4 units of the same structure,
and we choose one of the units to propagate our simulations at pH = 7 and 1.
The protein structures are generated by CHARMM-GUI:\cite{WonpilIm2008,WonpilIm2016}
At pH = 7, the protein carries 11 negative charges, which is neutralized by 11 Na$^+$ ions;
At pH = 1, the protein carries 35 positive charges, which is neutralized by 35 Cl$^-$ ions.
In both pH environments,
the protein is
solvated by a TIP3P water molecules in box of about 9.5 nm $\times$ 9.5 nm $\times$ 9.5 nm
to run relaxing simulations.
Both protein chains yield \ac{Rg}$\sim 2.2$ nm,
after running relaxation $NVT$ at 300 K for 10 ns,
$NPT$ at 300 K and 1 atm for 10 ns,
and another $NVT$ at 300 K for 10 ns.
At both pH = 7 and 1,
we simulate the \ac{Pdot} interacting with 6 different faces of protein,
by rotating the protein along the x-, y-, and z-axis as the 6 faces of the dice;
see Figures~\ref{fig:model} and~S4 in the \ac{SI}.

The binding interaction setup is similar to out previous work on
cytochrome C protein interacting with EG6-coated nanoparticles.\cite{hern20g}
The initial distance between the protein surface and the \ac{Pdot} surface was
set to $\sim2$ nm in Figure~\ref{fig:model}a.
The 6 amino acids with the lowest x-, y-, and z-coordinates,
and the highest x-, y-, and z-coordinates 
were labeled to identify the 6 different faces on the protein;
see Figure~\ref{fig:model}b and
Figure~S5 in the \ac{SI}.
A representative case in Figure~\ref{fig:model} shows
the scheme of face 1 at pH = 7,
where the residue number 128 PRO is facing the \ac{Pdot}.
The resulting final box size is
about 16.5 nm $\times$ 9.5 nm $\times$ 9.5 nm
by extending the box in the interacting direction,
and the total number of atoms is 152,218 for interacting face 1 at pH = 7;
see Figure~S5b in the \ac{SI}.
Similarly, Figure~S6 in the \ac{SI} shows that
at pH = 1, we use the 6 amino acids
to label the 6 different faces on the \ac{Pdot}.
As the two proteins are relaxed independently at pH = 7 and 1,
the protein conformations and the selected amino acids are different.
For comparison convenience,
we rotate the protein visualization angle at pH = 1
to make its labeled faces 1 to 6 approximately represent the same
binding directions with that at pH = 7 in Figures~S5 and~S6 in the \ac{SI}.

\subsection{Simulation Protocol}
The \ac{OPLS} all-atom force field is used to simulate the
atomistic interactions in \acp{Pdot}.\cite{jorgensen96, jorgensen01}
The CHARMM36 force field is used to simulate the ovalbumin protein.\cite{charmm36}
The cross \ac{LJ} interactions between \ac{Pdot} and protein are calculated
by the Lorentz-Berthelot mixing rule.
Initially, the \ac{Pdot} models at pH = 7 and 1 were built and relaxed using
the \ac{LAMMPS} package.\cite{plimpton95}
The protein models at pH = 7 and 1 are built by CHARMM-GUI and relaxed using the
Gromacs package.\cite{abraham2015gromacs}
After relaxing,
the \ac{Pdot} and protein models are combined resulting in 6 different interacting faces at each pH.
The final 12 systems of \ac{Pdot}-protein binding simulations are propagated
using the Gromacs package.\cite{abraham2015gromacs}
The NVIDIA A100 GPUs 
supported by the UIUC Delta HPC center through the NSF ACCESS proposal
were used to run all the Gromacs \ac{MD} simulations.
The code performance is about  20 ns/day for our \ac{Pdot}-protein systems with
a total number of about 152,000 atoms.
The simulation time step is set to 1~fs.
The cutoff distances are at 1.2~nm for both Coulombic and \ac{LJ} forces.
The initial systems are shown in Figure~\ref{fig:model} at pH = 7 and Figure~S4 in the \ac{SI} at pH = 1.
Typically, each system was relaxed under $NPT$ at 300~K and 1~atm for 2~ns,
followed by $NVT$ at 300~K for 2~ns,
and this relaxing process was repeated for more than 5 times
with a total of 20 ns $NPT$ and 20 ns $NVT$ relaxation.
Then, the productive run is performed by pulling the \ac{Pdot} towards the protein
under $NVT$ at 300~K for 90 ns, until two surfaces touch each other.
Representative trajectories
about the pulling process are shown in Figures~S7 and~S8.
The productive run is replicated by 5 times using different random seeds.
\subsection{PMF Calculation}
During the pulling process in the productive runs,
we dump a trajectory every 1 ns  to conduct umbrella sampling.
Applying the umbrella sampling method in \ac{MD} simulations to
calculate \acp{PMF} is well documented in literature.\cite{torrie1977,bernardi2015,tuckerman2010}
Briefly,
the unbiased probability distribution in each window, $i$, in the reaction coordinate, $\xi$, is
calculated as $P^{u}_{i}(\xi)$ in Eq.~\ref{eq:e1}.
\begin{equation}
P^{\text{u}}_{i}(\xi) = P^{\text{b}}_{i}(\xi) \text{exp}[\beta W_{i}(\xi)] \langle \text{exp}[-\beta W_{i}(\xi)] \rangle,
\label{eq:e1}
\end{equation}
where
$\xi$ is the \ac{COM}-\ac{COM} distance between \ac{Pdot} and protein,
$P^{\text{b}}_{i}(\xi)$ is the biased probability distribution obtained from \ac{MD} simulation trajectories,
and the harmonic potential bias is $W_{i}(\xi) = \frac{K}{2}(\xi - \xi_i^{\text{ref}})^2$
with $K = 1000$ kJ/mol/nm$^2$ set in our simulation.
The free energy \ac{PMF} curve $A(\xi)$ is related to $F_{i}$ by Eq.~\ref{eq:e2}.
\begin{equation}
\begin{split}
\text{exp}(-\beta F_{i}) &= \langle \text{exp}[-\beta W_{i}(\xi)] \rangle \\
&= \int  P^{\text{u}}(\xi) \text{exp}[-\beta W_{i}(\xi)] d\xi \\
&= \int  \text{exp}\{-\beta[A(\xi) - W_{i}(\xi)]\} d\xi,
\label{eq:e2}
\end{split}
\end{equation}
where $P^{\text{u}}(\xi)$ is the unbiased probability distribution.
The \ac{WHAM} is applied to compute the free energy $A(\xi)$.\cite{kumar1992,kastner2011}
This method starts with an initial guess of $F_{i} = 0$,
next it calculates $p_{i}(\xi)$ and $P^{\text{u}}(\xi)$ by Eqs.~\ref{eq:e3}--\ref{eq:e5},
then it calculates $A(\xi)$ from $P^{\text{u}}(\xi)$ and the new $F_{i}$,
and it continues this loop until converge.
The convergence tolerance is $F_{i}$ at $10^{-8}$.
%
\begin{equation}
P^{\text{u}}(\xi) = \sum_{i=1}^{N} p_{i}(\xi) P^{\text{u}}_{i}(\xi),
\label{eq:e3}
\end{equation}
\begin{equation}
\sum_{i=1}^{N} p_{i}(\xi) = 1,
\label{eq:e4}
\end{equation}
\begin{equation}
p_{i}(\xi) = \frac{n_{i} \text{exp}[-\beta W_{i}(\xi) + \beta F_{i}]}{\sum_{j=1}^{N} n_{j} \text{exp}[-\beta W_{j}(\xi) + \beta F_{j}]}.
\label{eq:e5}
\end{equation}

\begin{figure}[t]
\includegraphics[clip=true,scale=0.25,width=1\linewidth]{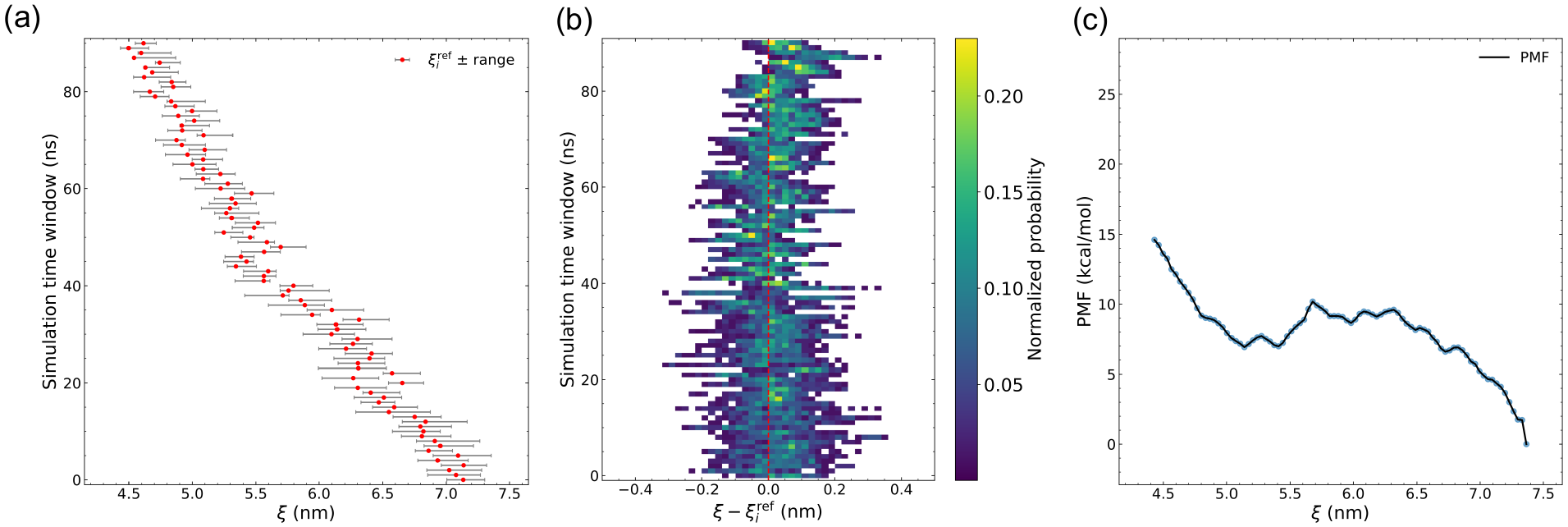}
\caption{
An example of PMF calculation using WHAM for face 1 at pH = 7.
(a) The trajectories are saved every 1 ns shown by the red dots for the 90 ns pulling process,
where the reaction coordinate, $\xi$, is the \ac{COM}-\ac{COM} distance of \ac{Pdot} and protein.
For each trajectory,
we perform 1 ns umbrellas sampling simulation, using a harmonic bias potential,
and the error bars show the upper and lower bounds of $\xi$ in the sampling window.
(b) The distribution of $\xi$ in each 1 ns sampling trajectories with 100 frames.
(c) The PMF calculated by Eqs~\ref{eq:e1}-\ref{eq:e5} using the distributions in (b).
}
\label{fig:method_pmf}
\end{figure}

Figure~\ref{fig:method_pmf} shows a representative case about \ac{PMF} calculation in detail.
We pull the \ac{Pdot} towards the protein along the reaction coordinate for 90 ns,
where the intermediate structures are saved every 1 ns; see Figure~\ref{fig:method_pmf}a.
Using these saved structures,
we apply a harmonic bias potential between the protein and the \ac{Pdot},
which allows us to sample many trajectories near that position.
Typically, we perform a 1 ns sampling simulation time and obtain 100 frames in each window
using a dumping frequency of 10 ps per frame.
The distributions from these sampling windows have overlaps as shown in Figure~\ref{fig:method_pmf}b.
The \ac{PMF} curve is calculated
in Figure~\ref{fig:method_pmf}c by setting the farthest distance to PMF = 0.
In addition,
preliminary results show that
multiple replicated runs can well cover the uncertainty of \ac{PMF} calculations,
which is better than increasing the sampling simulation time from 1 ns to 10 ns;
see Figures~S9 to S11 in the \ac{SI}.
For each \ac{Pdot}-protein interacting face,
we performed 5 independent umbrellas sampling simulations for \ac{PMF} calculation
and reported the uncertainty.

\section{Results and Discussion}

\subsection{Electrostatic and \ac{LJ} interaction energies of Pdot binding to protein at pH = 7}

\begin{figure}[t]
\includegraphics[clip=true,scale=0.25,width=0.8\linewidth]{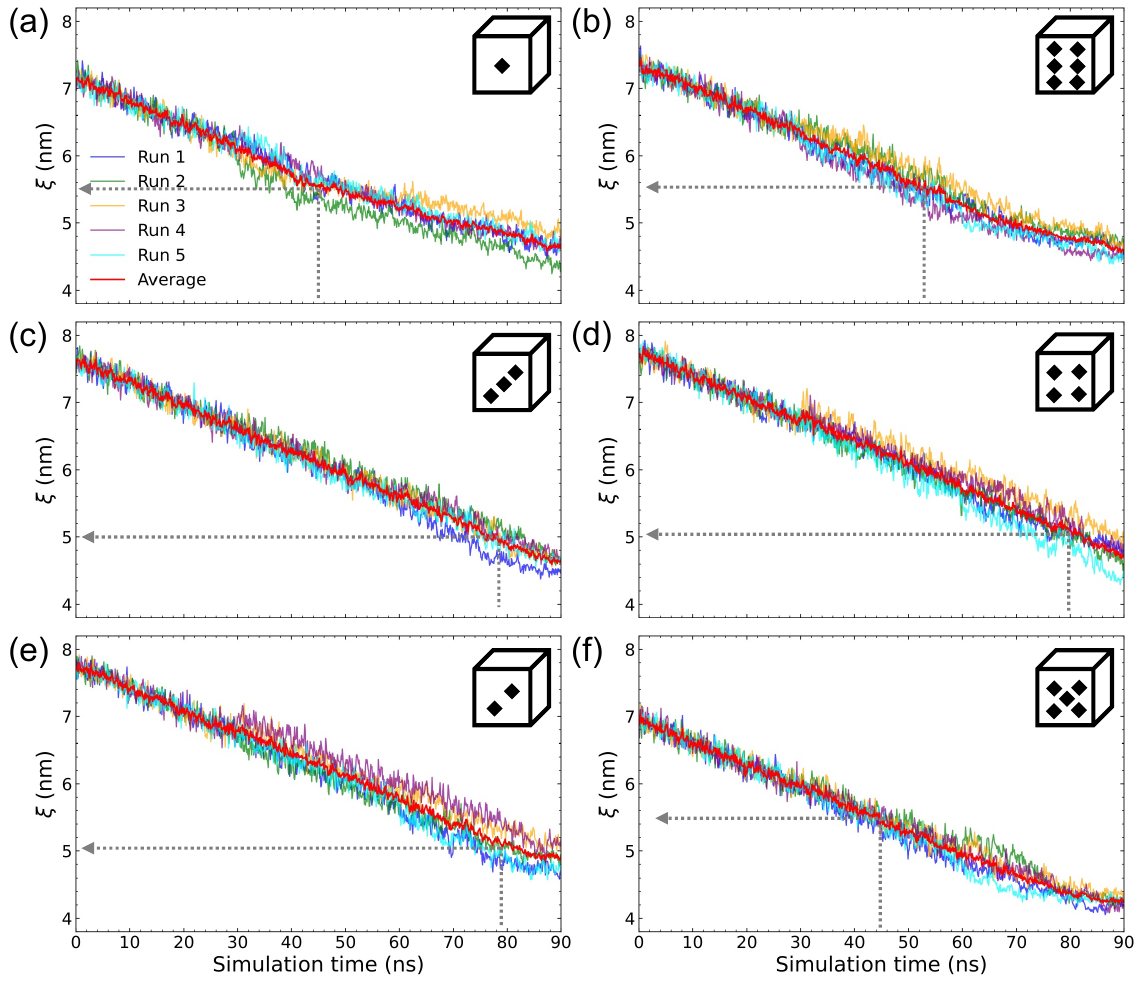}
\caption{
(a) - (f) The reaction coordinate, $\xi$, changing
during the 90 ns pulling process
for the 6 different interacting faces at pH = 7 as indicated in Figure~\ref{fig:model}.
In each case, the 5 independent runs and their averaged value are shown.
The gray dashed lines show the approximate $\xi$,
when the interactions happen using times---grey dashed lines---in Figure~\ref{fig:energy_pH7}.
}
\label{fig:distance_pH7}
\end{figure}

\begin{figure}[t]
\includegraphics[clip=true,scale=0.25,width=0.8\linewidth]{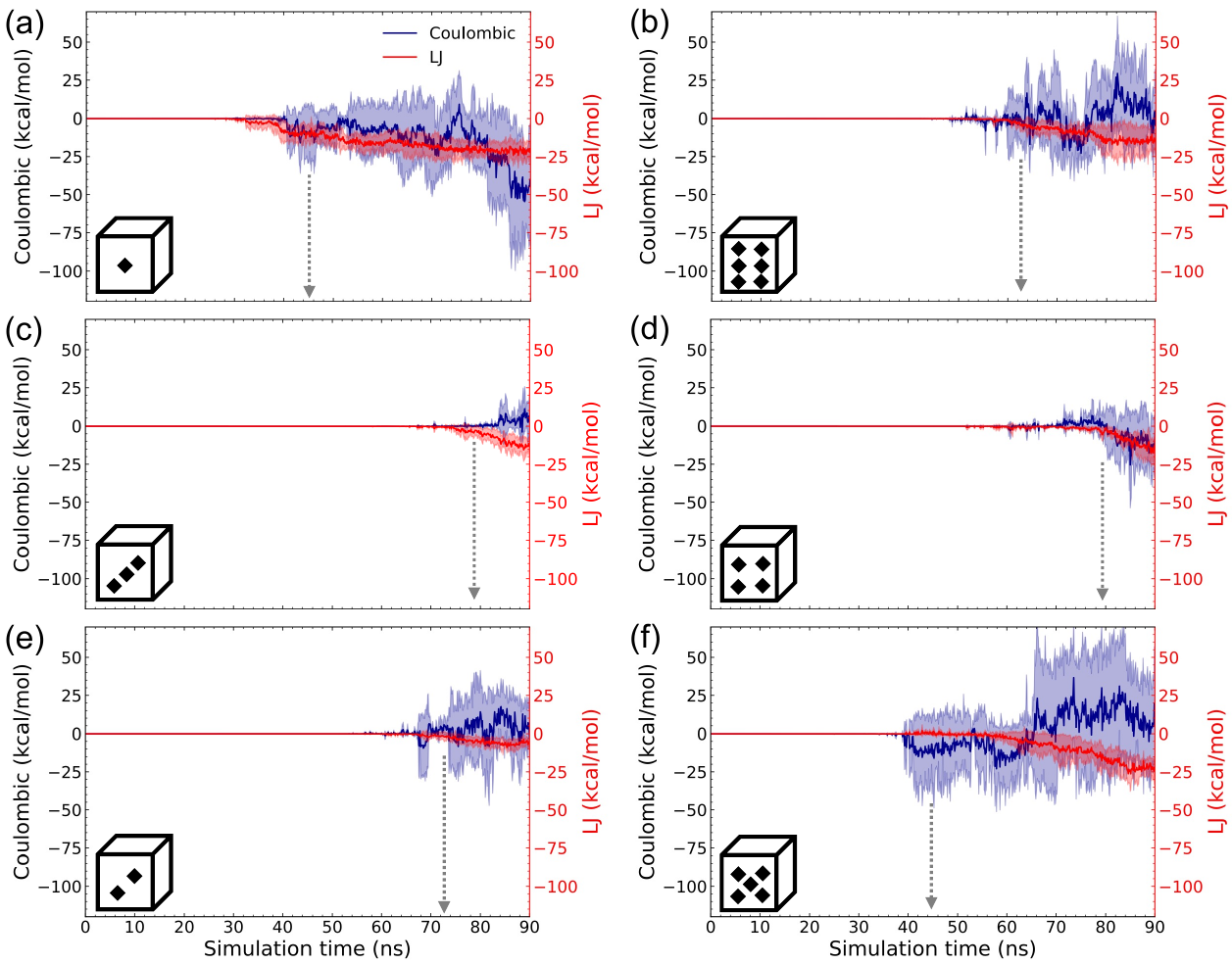}
\caption{
(a) - (f) The Coulombic in blue and LJ in red interaction energies changing
during the 90 ns pulling process
for the 6 different interacting faces at pH = 7 as indicated in Figure~\ref{fig:model}.
In each case, the averaged value from 5 independent runs are plotted
together with the standard deviation in shaded color.
The gray dashed lines show the approximate simulation time, when the interacting starts.
}
\label{fig:energy_pH7}
\end{figure}

As we set the initial distance between the \ac{Pdot} surface and the protein surface at about 2 nm,
and the \ac{Rg} values for both \ac{Pdot} and protein are 2 - 3 nm,
resulting at $t=0$ ns the \ac{COM}-\ac{COM} distance $\xi=$ 7 - 8 nm  for all different interacting faces
in Figure~\ref{fig:distance_pH7}.
At distance setting,
the Electrostatic and \ac{LJ} interaction energies are zero in Figure~\ref{fig:energy_pH7} at $t=0$ ns.
Figure~\ref{fig:distance_pH7} shows that 5 independent pulling trajectories
are performed using different random seeds to replicate the binding process.
For example, in the face 1 case in Figure~\ref{fig:distance_pH7}a,
the protein is aligned with the \ac{Pdot} in the x-axis,
resulting the x-component of $\xi$ is dominant in Figure~S12 in the \ac{SI}
when pulling the \ac{Pdot} to the protein.
We show that
the slope of $\xi$ in Figure~\ref{fig:distance_pH7} is the pulling speed at $\sim0.03$ nm/ns,
which is slow enough for sampling trajectories.
In steered \ac{MD} simulations, the typical pulling speed is about 1 nm/ns.\cite{hern23c,hern24d,hern26b}
We also observe that after the protein touches the \ac{Pdot},
the pulling speed is slower;
see the slopes are reduced at 50 - 90 ns in Figure~\ref{fig:distance_pH7}a
and at 70 - 90 ns in Figure~\ref{fig:distance_pH7}b.
Using the Coulombic and \ac{LJ} interaction energies in Figure~\ref{fig:energy_pH7},
we can approximately identify the simulation times when the \ac{Pdot} start to interact with the protein.
Due to the heterogeneity of the model,
for faces 1 and 5, the interactions start after 40 ns shown in
Figures~\ref{fig:distance_pH7}a and~\ref{fig:distance_pH7}f, respectively;
while for faces 3 and 4, the interactions start after 70 ns shown in
Figures~\ref{fig:distance_pH7}c and~\ref{fig:distance_pH7}d, respectively.
The representative all 5 independent runs in Figure~S13 in the \ac{SI}
shows that the uncertainty in Coulombic interaction energy is much larger than \ac{LJ} interaction energy.

The binding affinity between the \ac{Pdot} and the protein is related to the simulation time
before the interactions starts not after.
For example, in Figure~\ref{fig:distance_pH7}a face 1,
the initial interacting stage starts near 45 ns,
after that the \ac{Pdot} and protein are squeezed. 
The protein \ac{RMSD} changing in Figure~S14 in the SI
and protein \ac{Rg} changing in Figure~S15 in the SI
help track the protein structure changing during the pulling process,
which shows that in general, the \ac{RMSD} and \ac{Rg} seem stay similar for the 6 cases,
but the \ac{Rg} in x-direction for face 1 in Figure~S15a can correspond to
the squeezing after 45 ns in Figure~\ref{fig:distance_pH7}a.
Figure~\ref{fig:distance_pH7} also shows that
at the initial contact stage,
the \ac{LJ} interaction energies are similar for all 6 faces,
but the Coulombic interaction energies vary significantly.
The small negative \ac{LJ} interaction energy of $\sim-10$ kcal/mol
means that a weak \ac{vdW} force exists between \ac{Pdot} and protein in all 6 faces.
A strong fluctuation in the Coulombic interaction energy of $\pm25$ kcal/mol
indicates that the electrostatic interaction is strong,
which is related to the charge distribution on the protein.
Figures~\ref{fig:distance_pH7}a, \ref{fig:distance_pH7}d, and \ref{fig:distance_pH7}e
give negative Coulombic interaction energies,
which may suggest that
faces 1, 4, and 5 have higher binding affinity than the other faces at pH = 7.
In the next section, we use the \ac{PMF} calculations to validate the above hypothesis.

\subsection{PMF of Pdot binding to protein at pH = 7}

\begin{figure}[t]
\includegraphics[clip=true,scale=0.25,width=0.8\linewidth]{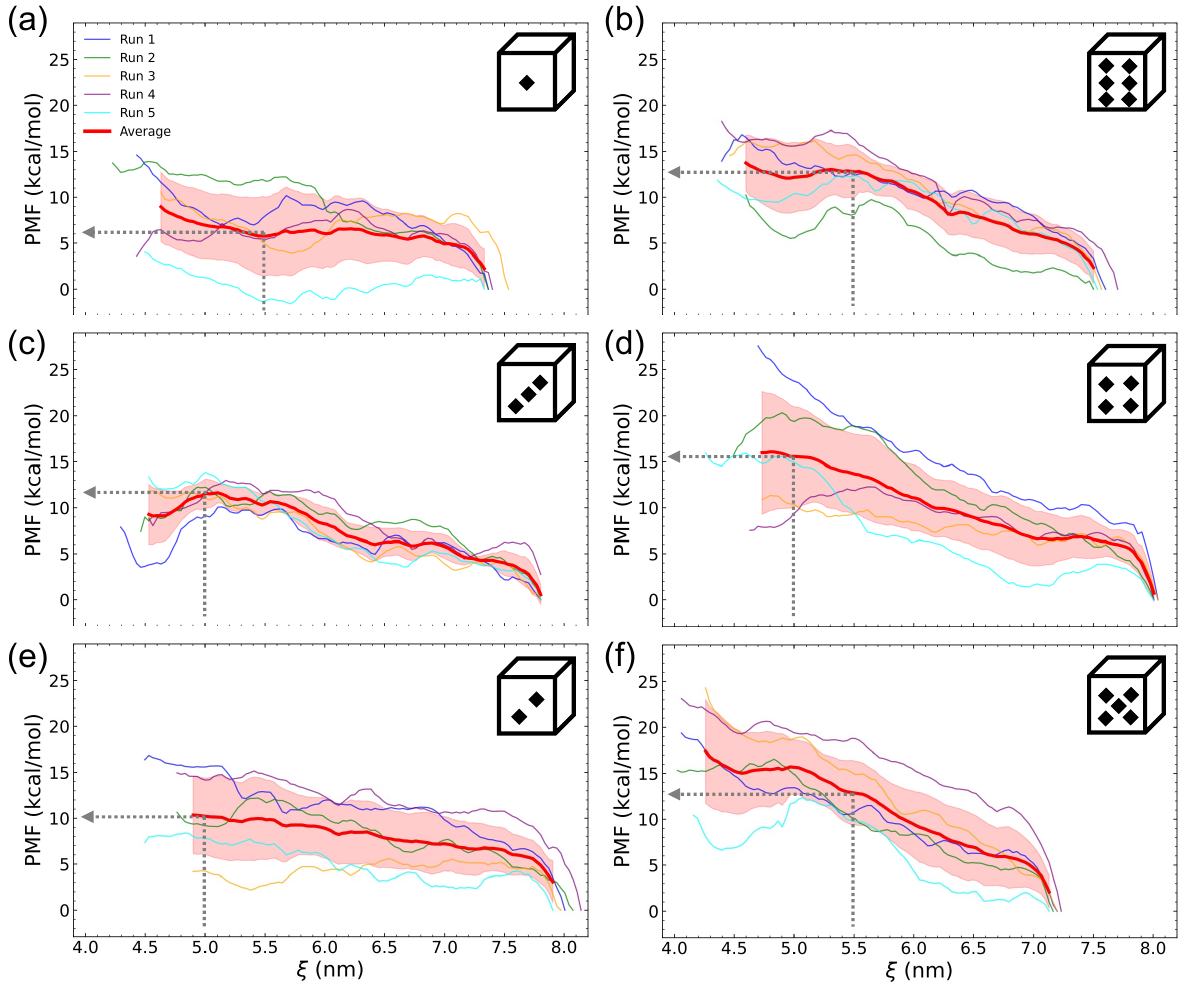}
\caption{
(a) - (f) The \ac{PMF} along the interaction coordinate, $\xi$,
in the 6 different interacting faces indicated in Figure~\ref{fig:model} for pH = 7 models.
In each panel, all 5 independent runs are shown together with the average in red line
and standard deviation in shaded area.
The gray dashed lines show the corresponding PMF,
using the $\xi$ in Figure~\ref{fig:distance_pH7}.
}
\label{fig:PMF_pH7}
\end{figure}

Figure~\ref{fig:PMF_pH7} shows the \ac{PMF} curves
that we calculated for the 6 different interacting faces at pH = 7.
We find that
all 6 interacting faces lead to increasing \ac{PMF},
when the \ac{Pdot} approaches the protein at pH = 7.
Table~\ref{tab:PMF_xi} summarizes the approximate interacting times, distances, and \ac{PMF} values.
For example,
for face 1 at pH = 7,
we find the interaction approximately starts at 45 ns in Figure~\ref{fig:energy_pH7}a,
which leads to $\xi \sim 5.5$ nm using $t=45$ ns in Figure~\ref{fig:distance_pH7}a,
resulting PMF$\sim 6$ kcal/mol using $\xi \sim 5.5$ nm in Figure~\ref{fig:PMF_pH7}a.
Similarly, for faces 6 and 5 at pH = 7,
the initial interacting distances are both at $\xi \sim 5.5$ nm
in Figures~\ref{fig:distance_pH7} and~\ref{fig:energy_pH7},
using $t=45$ and 53 ns, respectively,
and resulting PMF$=10$ to 15  kcal/mol in Figures~\ref{fig:PMF_pH7}b and \ref{fig:PMF_pH7}f, respectively.
For faces 3, 4, and 2 at pH = 7,
we find that
the initial interacting time is near 80 ns where $\xi \sim 5.0$ nm
and resulting PMF $\sim$ 12, 16, and 10 kcal/mol, respectively;
see the corresponding panels in Figures~\ref{fig:distance_pH7} to \ref{fig:PMF_pH7}.
In the previous section,
the Coulombic and \ac{LJ} interaction energies in Figure~\ref{fig:energy_pH7}
suggest that faces 1, 4, and 5 may have better binding affinity than other faces,
but this is partially disproved by the PMF results in Figure~\ref{fig:PMF_pH7}.
Because we find that faces 4 and 5 have the large PMF at about 16 and 13 kcal/mol, respectively;
see Figure~\ref{fig:PMF_pH7}.
Meanwhile, face 1 is consistent with the above hypothesis,
which gives the lowest PMF of about 6 kcal/mol.
However, all PMF values have positive numbers at pH = 7,
which suggest that
the \ac{Pdot} does not have good binding affinity with the protein.
\begin{table}[ht]
\centering
\caption{Summary of when the interactions start, the approximated simulation time,
the interacting distance ($\xi$), and the corresponding PMF,
for different faces at pH = 7 and 1.}
\label{tab:PMF_xi}
\begin{tabular}{c|cccc}
\toprule
 & Face & Time (ns) & $\xi$ (nm) & PMF (kcal/mol) \\
\midrule
\multirow{6}{*}{pH = 7}
& 1 & 45 & 5.5 & 6 \\
& 2 & 73 & 5.0 & 10 \\
& 3 & 79 & 5.0 & 12 \\
& 4 & 80 & 5.0 & 16 \\
& 5 & 45 & 5.5 & 13 \\
& 6 & 63 & 5.5 & 13 \\
\midrule
\multirow{6}{*}{pH = 1}
& 1 & 68 & 5.7 & -5 \\
& 2 & 62 & 4.8 & 3 \\
& 3 & 52 & 4.8 & 5 \\
& 4 & 58 & 4.8 & 8 \\
& 5 & 58 & 5.2 & 17 \\
& 6 & 84 & 5.4 & -30 \\
\bottomrule
\end{tabular}
\end{table}

\subsection{Electrostatic and \ac{LJ} interaction energies of Pdot binding to protein at pH = 1}

\begin{figure}[t]
\includegraphics[clip=true,scale=0.25,width=0.8\linewidth]{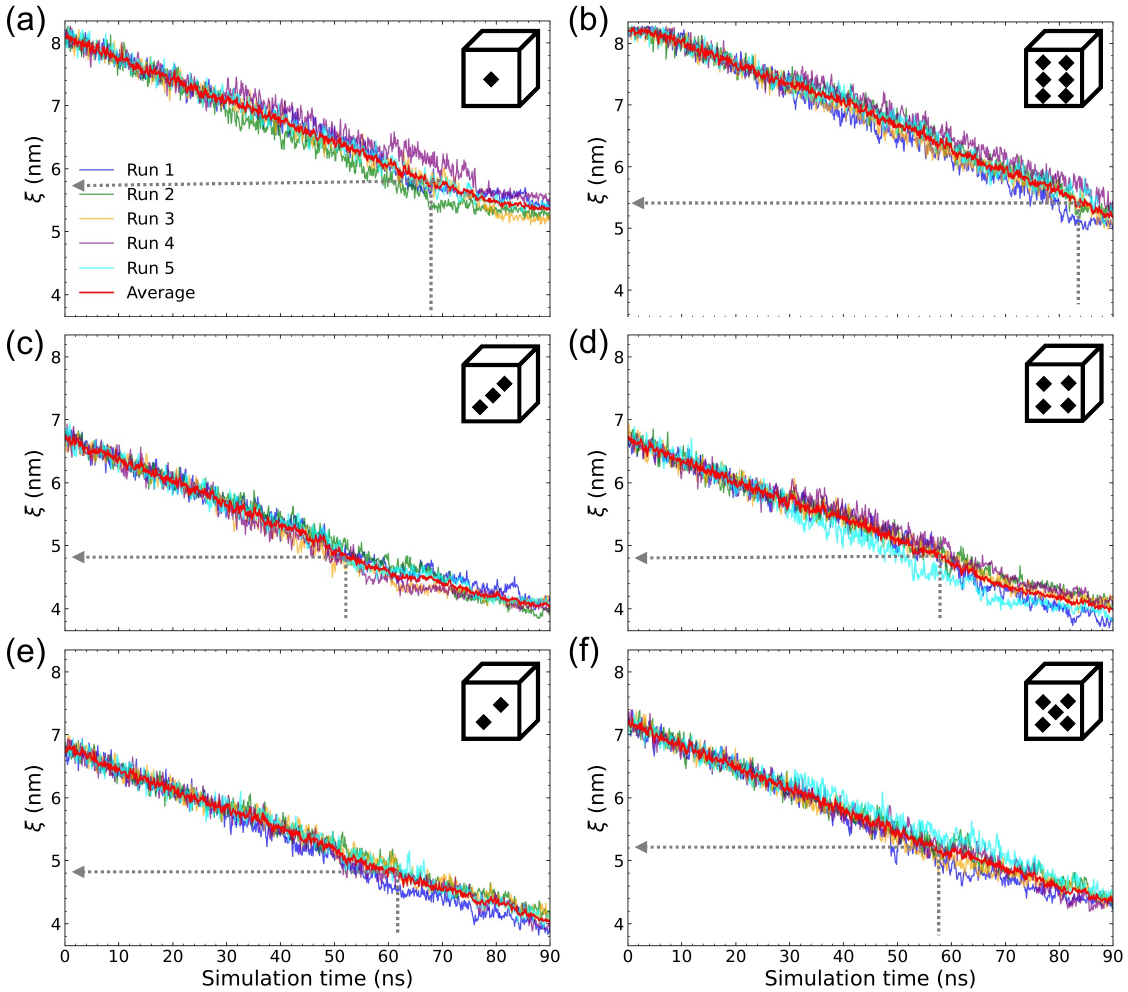}
\caption{
(a) - (f) The reaction coordinate, $\xi$, changing in the 90 ns pulling process for the 6 different
interacting faces at pH = 1 as indicated in Figure~S4 in the SI.
In each case, the 5 independent runs and their averaged value are shown.
The gray dashed lines show the approximate $\xi$,
when the interactions happen using times---grey dashed lines---in Figure~\ref{fig:energy_pH1}.
}
\label{fig:distance_pH1}
\end{figure}

\begin{figure}[t]
\includegraphics[clip=true,scale=0.25,width=0.8\linewidth]{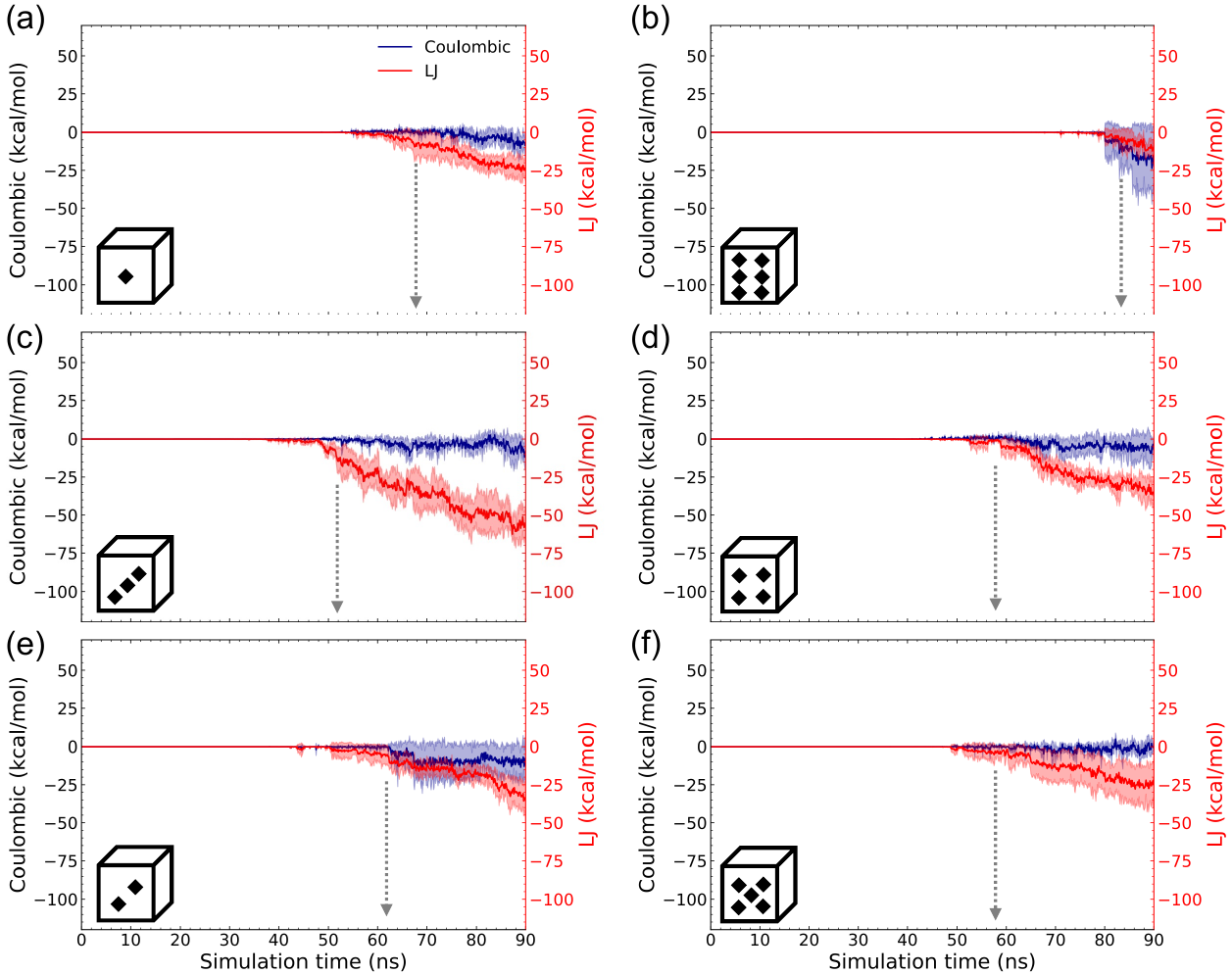}
\caption{
(a) - (f) The Coulombic in blue and \ac{LJ} in red interaction energies changing
during the 90 ns pulling process
for the 6 different interacting faces at pH = 1 as indicated in Figure~S4 in the SI.
In each panel, the average of 5 independent runs are plotted
together with the standard deviation in shaded color.
The gray dashed lines show the approximate simulation time, when the interacting starts.
}
\label{fig:energy_pH1}
\end{figure}

Similarly, we investigate the binding affinities for 6 different faces on the protein at pH = 1
following the same protocol as that at pH = 7.
Figure~\ref{fig:distance_pH1} shows the 5 independent trajectories in each case 
for the 6 different interacting faces,
where slopes of the trajectories give the pulling speed of $\sim0.03$ nm/ns
before the \ac{Pdot} and the protein touches,
and after that the pulling speed decreases.
In general,
we did not observe the protein structure changing during the pulling process;
see the \ac{RMSD} in Figure~S16 and
the \ac{Rg} in Figure~S17.
The approximate simulation times when the \ac{Pdot} start contacting the protein
can be found using the Coulombic and \ac{LJ} interaction energies in Figure~\ref{fig:energy_pH1},
which gives
for faces 1, 6, 3, 4, 2, and 5, the approximate interacting times are 68, 84, 52, 58, 62, and 58 ns, respectively.
Using Figure~\ref{fig:distance_pH1},
we find the interacting distances for
face 1 is $\xi \sim$ 5.7 nm,
face 6 is $\xi \sim$ 5.4 nm,
face 3 is $\xi \sim$ 4.8 nm,
face 4 is $\xi \sim$ 4.8 nm,
face 2 is $\xi \sim$ 4.8 nm, and
face 5 is $\xi \sim$ 5.2 nm;
see Table~\ref{tab:PMF_xi} for the pH = 1 results.
We also find that for faces 1, 3, 4, and 5 at pH = 1,
the Coulombic interaction energies have
magnitudes near 0,
which is much smaller than that at pH = 7.
For faces 6 and 2,
in Figures~\ref{fig:energy_pH1}b and~\ref{fig:energy_pH1}e, respectively,
the Coulombic interaction energies are about -20 and -10 kcal/mol,
which means better binding affinity.
When the \ac{Pdot} is pushed closer into the protein,
the \ac{LJ} interaction energies are lower than -25 kcal/mol at 90 ns;
see Figures~\ref{fig:energy_pH1}c - \ref{fig:energy_pH1}e.
However, at the approximated interaction starting time listed in Table~\ref{tab:PMF_xi},
we find the magnitudes of \ac{LJ} interaction energies are similar to
all cases at pH = 7---less than 10 kcal/mol.
This is because the \ac{LJ} energies are calculated based on the nearest neighbor atoms,
which is generally similar in all cases;
when the \ac{Pdot} and the protein are pushed closer,
the number of nearest neighbors increases and the \ac{LJ} interaction energy increases.
In all,
the Coulombic interaction energies in Figure~\ref{fig:energy_pH1} suggest that
faces 6 and 2 at pH = 1 have the best binding affinities than other faces.
In the following section,
we calculate the \ac{PMF} at pH = 1 to verify the above hypothesis.

\subsection{PMF of Pdot binding to protein at pH = 1}

\begin{figure}[t]
\includegraphics[clip=true,scale=0.25,width=0.8\linewidth]{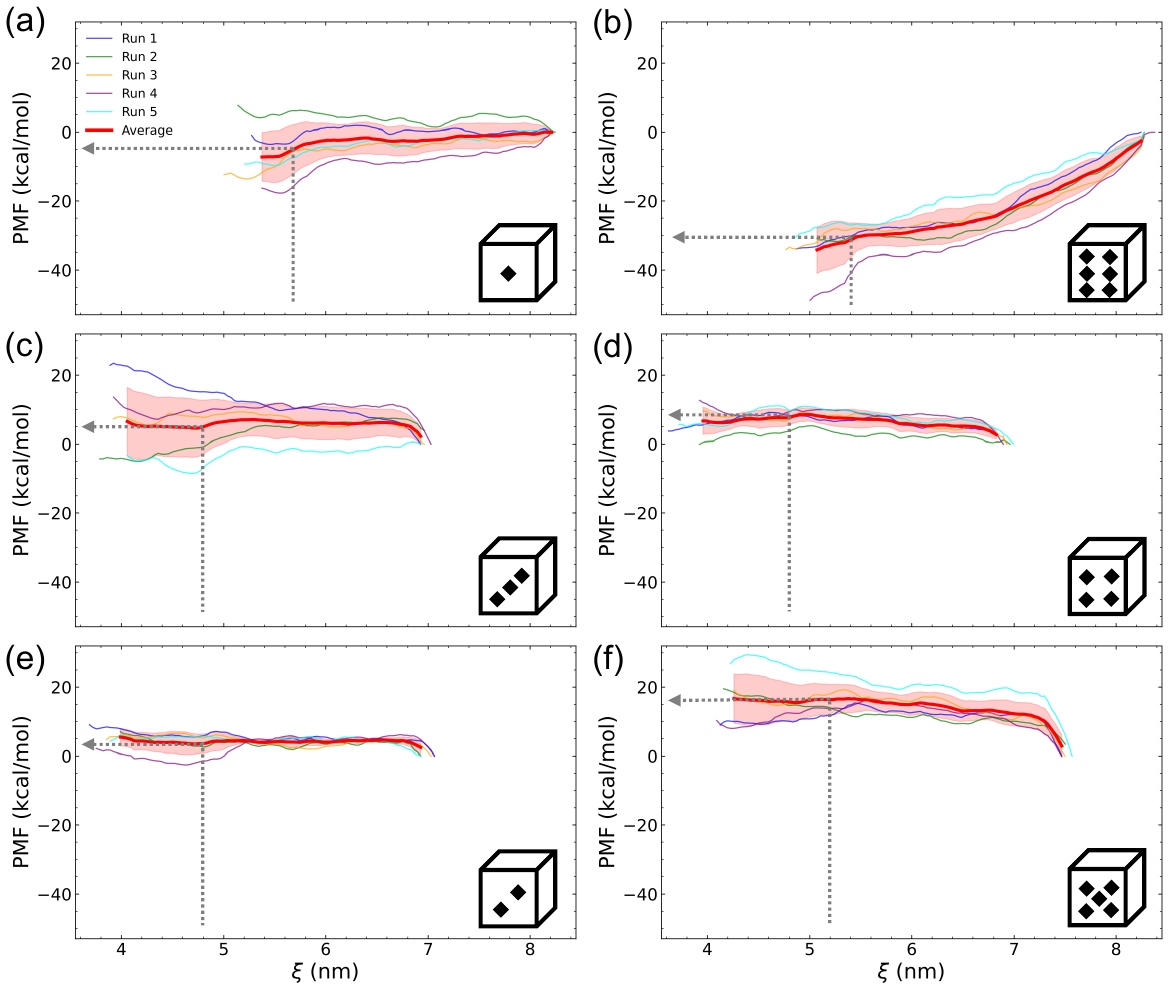}
\caption{
(a) - (f) The PMF along the interaction coordinate, $\xi$,
for the 6 different interacting faces at pH = 1 as indicated in Figure~S4 in the SI.
In each panel, all 5 independent runs are shown together with the average in red line
and standard deviation in shaded area.
The gray dashed lines show the corresponding PMF,
using the $\xi$ in Figure~\ref{fig:distance_pH1}.
}
\label{fig:PMF_pH1}
\end{figure}

Figure~\ref{fig:PMF_pH1} show the \ac{PMF} curves
for the 6 different interacting faces at pH = 1,
which yield both positive and negative values.
In Table~\ref{tab:PMF_xi} for pH = 1, we show that
using the initial interacting times and distances,
we can determine the \ac{PMF} values for 
face 1 is about $-5$ kcal/mol at $\xi \sim$ 5.7 nm,
face 6 is about $-30$ kcal/mol at $\xi \sim$ 5.4 nm,
face 3 is about 5 kcal/mol at $\xi \sim$ 4.8 nm,
face 4 is about 8 kcal/mol at $\xi \sim$ 4.8 nm,
face 2 is about 3 kcal/mol at $\xi \sim$ 4.8 nm, and 
face 5 is about 17 kcal/mol at $\xi \sim$ 5.2 nm.
It is important that
face 6 in Figure~\ref{fig:PMF_pH1}b has the lowest \ac{PMF} of about $-30$ kcal/mol,
which is consistent with the hypothesis
based on the Coulombic interaction energy in Figure~\ref{fig:energy_pH1}b
in the previous section.
On the other hand,
face 2 has a \ac{PMF} value of about 3 kcal/mol
meaning the binding affinity is not strong, but ,
which can not support the hypothesis in the previous section.
Meanwhile, face 1 also shows a small negative \ac{PMF}$ = -5$ kcal/mol, meaning good binding affinity.
Faces 3 and 4 have \ac{PMF} of 5 and 8 kcal/mol,
which is larger than face 2,
but much smaller than face 5 with \ac{PMF} about 17 kcal/mol.
Comparing to the 6 faces at pH = 7 in Table~\ref{tab:PMF_xi},
we find that the \ac{PMF} values at pH = 1 are much smaller in general,
except the face 5 in Figure~\ref{fig:PMF_pH1}f.
Meanwhile,
using the \ac{PMF} $\sim-30$ kcal/mol for face 6  in Figure~\ref{fig:energy_pH1}b,
we have strongly demonstrated that
the \ac{Pdot} can bind to the protein at pH = 1 better than at pH = 7. 
Our simulation results agree with the experimental observation reported by \citet{zeev25}
showing that decrease pH can increase the number of proteins binding to the \ac{Pdot}.

\subsection{Charge distribution on the protein}

\begin{figure}[t]
\includegraphics[clip=true,scale=0.25,width=0.8\linewidth]{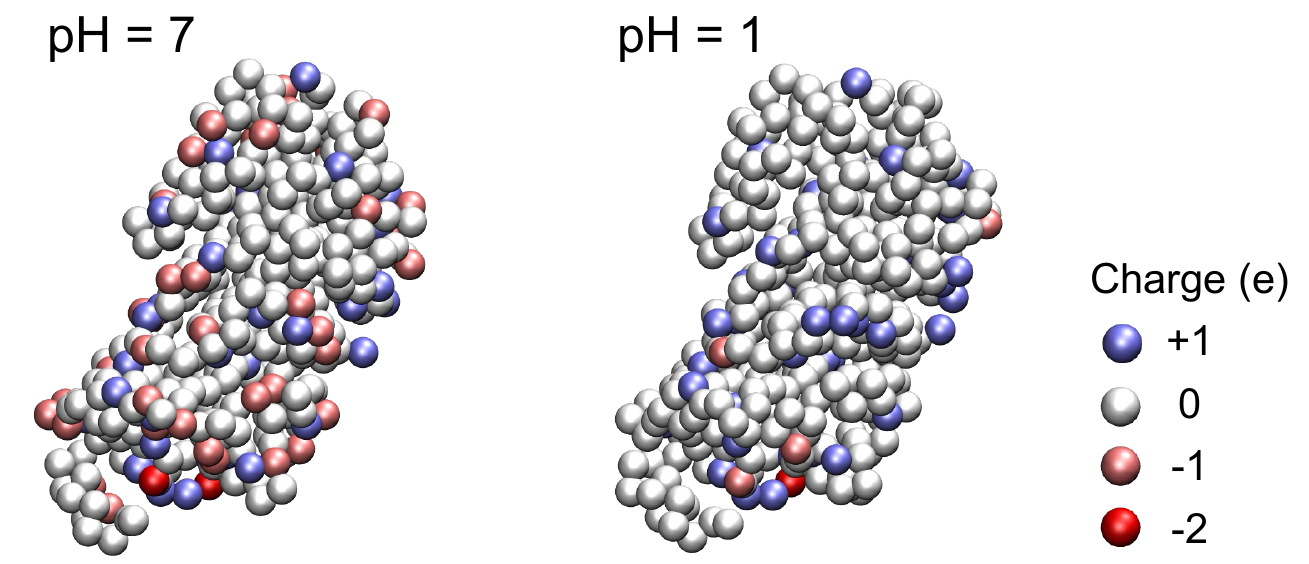}
\caption{
The residue charge distribution of the protein at pH = 7 and 1,
for the conformations of face 6 in
Figures~\ref{fig:model} and S4 in the \ac{SI}, respectively.
Each bead represents the alpha carbon of the residue.
Deep red, light red, white, and light blue colors mean
the residue has a net charge of -2, -1, 0, and +1 e, respectively.
No reside has +2 e charges in both cases.
The total net charge at pH = 7 is $-11$ e and at pH = 1 is $+35$ e.
}
\label{fig:charge}
\end{figure}

At pH = 7, the ovalbumin protein carries $-11$ e net charges, and
the \ac{PFBT} part on the \ac{Pdot} with COO$^-$ groups also carries negative charges.
In general,
the electrostatic interactions between the \ac{Pdot} and the protein are repulsive,
which should result in a net positive Coulombic interaction energy and bad binding affinity.
The \ac{PMF} in Figure~\ref{fig:PMF_pH7}
shows that all 6 interacting faces have large positive \ac{PMF} values,
which can confirm
weak binding affinity between the \ac{Pdot} and protein at pH = 7.
Figure~\ref{fig:energy_pH7} also show that at pH = 7, the Coulombic interaction energies
can fluctuating between positive and negative values significantly,
which is because the charge distribution on the protein is non-uniform,
and positively and negatively charged atoms are mixed.
Figure~\ref{fig:charge} shows that at pH = 7,
the negatively charged residues dominate on the protein surface,
but a large number of positively charged residues coexist,
which leads to the Coulombic interaction energy with high uncertainty.

Meanwhile, at pH = 1, the protein carries $+35$ e net charges
and the \ac{Pdot} is charge neutral with COOH groups,
which leads to near zero Coulombic interaction energies
in panels (c), (d), and (f) in Figure~\ref{fig:energy_pH7}
corresponding to faces 3, 4, and 5, respectively.
Figure~\ref{fig:charge} shows that at pH = 1 negative charges are neutralized
and the protein surface is dominated by positively charged residues.
The face 6 at pH = 1 has the strongest binding affinity---%
viz. the most negative Coulombic interaction energy and the lowest \ac{PMF}.
As the protein structure is heterogeneous and the charge distribution is non-uniform,
the face 6 has much better binding affinity than other faces.
The representative structures of \acp{Pdot} binding to the protein face 6 at both pH = 7 and 1 are
provided in Figure~S18 in the \ac{SI},
which shows that the actual contact areas on face 6 are not exactly the same
because the relaxing simulations are performed indepenently.
At pH = 7, all 6 faces have positive \ac{PMF} and bad binding affinity,
while at pH = 1 both faces 1 and 6 have negative \ac{PMF}.
Particularly, the face 6 at pH = 1 has the best binding affinity
of \ac{PMF}$\sim -30$ kcal/mol.
In all, the protein residue charge distributions at pH = 7 and 1 confirm that
at pH = 7, the protein with negative net charges shows weak binding affinity with the negatively charged \ac{Pdot},
and at pH = 1, the protein with positive net charges shows strong binding affinity with the charge neutral \ac{Pdot}.

\FloatBarrier

\section{Conclusions}

In this work, we use all-atom \ac{MD} simulations 
and umbrella sampling to determine
the binding affinity between \ac{Pdot}
and ovalbumin protein at pH = 7 and 1.
The Coulombic and \ac{LJ} interaction energies are calculated and compared
for the pulling trajectory 
between 
the \ac{PFBT} side of \ac{Pdot} 
and the protein surface.
We performed the pulling process for 6 different initial
faces on the protein,
and each interacting face was replicated 5 times to obtain the uncertainty.
Using umbrella sampling trajectories and the \ac{WHAM},
the \ac{PMF} values for the same 6 different initial faces were also 
obtained.
We found that at pH = 7, all 6 faces yield positive \ac{PMF} values,
and hence resulted in bad binding affinity.
One of the faces resulted in
the lowest Coulombic interaction energy
and the lowest \ac{PMF} of about 6 kcal/mol.
We further found that
at pH = 1, the Coulombic interaction energies 
and the \ac{PMF} values are generally much smaller
than that at pH = 7.
At pH = 1, the face with the lowest Coulombic interaction energy
also yields a \ac{PMF} of about $-30$ kcal/mol,
consistent with excellent binding affinity.
We also use the protein residue charge distribution schemes to observe that:
(1) at pH = 7, the negatively charged residues dominate the protein surface
behavior,
(2) at pH = 1, the protein surface has more positively charged resides,
and 
(3) the heterogeneity in the residue charge distribution is the
origin of the differing
Coulombic interactions and \ac{PMF} values resulting
from  different initial faces.
In summary, we have found that 
the binding affinity of the \ac{Pdot} and the ovalbumin protein
can be improved by lowering the pH
as it primarily enhances the attractive electrostatic interactions.

\section*{Acknowledgments}
\label{sec:Acknowledgments}

This work has been partially supported by the 
National Science Foundation (NSF) through Grant No.~CHE 2102455.
The computing resources necessary for this work were
performed in part on Expanse 
at the San Diego Supercomputing Center
through allocation CTS090079 provided 
by \ac{ACCESS}, which is supported by National Science Foundation (NSF)
grants \#2138259, \#2138286, \#2138307, \#2137603, and \#2138296.
Additional computing resources
were provided by the Advanced Research Computing at Hopkins (ARCH) 
high-performance computing (HPC) facilities supported 
by the NSF MRI Grant (OAC-1920103).

\section*{Supplementary Information}
\label{sec:SI}

The Supporting Information (SI) is available free of charge at XX.
Figures~S1 and~S2 show the \ac{Pdot} models at pH = 7 and 1, respectively.
Figure~S3 shows the protein models at pH = 7 and 1.
Figures~S4 to~S11 show simulation models for the 6 difference faces at pH = 7 and 1,
and the \ac{PMF} calculation methods.
Figures~S12 to~S15 show representative
comparisons of \ac{COM}-\ac{COM} in x-, y-, and z-directions,
Coulombic and \ac{LJ} interaction energies in all 5 independent runs,
and the protein \ac{RMSD} and \ac{Rg} changing at pH = 7.
Figures~S16 and~S17 show the protein \ac{RMSD} and \ac{Rg} changing at pH = 1.
Figure~S18 shows the representative trajectories of \acp{Pdot} binding to the protein at pH = 7 and 1


\section*{Data Availability}

The data that support the findings of this work are available from 
the corresponding author upon reasonable request.


\FloatBarrier


\newcommand{\doi}[1]{\href{http://dx.doi.org/#1}{\nolinkurl{#1}}}
\bibliography{paper_Pdot.bib}

\end{document}